\documentclass[showpacs,10pt,twocolumn,prb]{revtex4-1}

\usepackage{amsmath}
\usepackage{amssymb}
\usepackage{graphicx}
\usepackage{amssymb}
\usepackage{graphics}
\usepackage{epsfig}
\usepackage{CJK}
\usepackage{color}

\begin{document}

\begin{CJK*}{GBK}{Song}
\title{Magnetic anisotropy and entropy change in trigonal Cr$_5$Te$_8$}
\author{Yu Liu,$^{1}$ Milinda Abeykoon,$^{2}$ Eli Stavitski,$^{2}$ Klaus Attenkofer,$^{2}$ and C. Petrovic$^{1}$}
\affiliation{$^{1}$Condensed Matter Physics and Materials Science Department, Brookhaven National Laboratory, Upton, New York 11973, USA\\
$^{2}$National Synchrotron Light Source II, Brookhaven National Laboratory, Upton, New York 11973, USA}
\date{\today}

\begin{abstract}
We present a comprehensive investigation on anisotropic magnetic and magnetocaloric properties of the quasi-two-dimensional weak itinerant ferromagnet trigonal Cr$_5$Te$_8$ single crystals. Magnetic-anisotropy-induced satellite transition $T^*$ is observed at low fields applied parallel to the $ab$ plane below the Curie temperature ($T_c$ $\sim$ 230 K). $T^*$ is featured by an anomalous magnetization downturn, similar to that in structurally related CrI$_3$, which possibly stems from an in-plane antiferromagnetic alignment induced by out-of-plane ferromagnetic spins canting. Magnetocrystalline anisotropy is also reflected in magnetic entropy change $\Delta S_M(T,H)$ and relative cooling power RCP. Given the high Curie temperature, Cr$_5$Te$_8$ crystals are materials of interest for nanofabrication in basic science and applied technology.
\end{abstract}
\maketitle
\end{CJK*}

\section{INTRODUCTION}

Achieving long-range magnetism at low dimensions as well as high temperature is of great importance for the development of next-generation spintronics. As stated in the Mermin-Wagner theorem,\cite{Mermin} materials with isotropic Heisenberg interactions can not magnetically order in the two-dimensional (2D) limit. Strong magnetic anisotropy such as exchange anisotropy and/or single-ion anisotropy may remove this restriction. Recently, 2D long-range ferromagnetism was observed in van der Waals (vdW) CrI$_3$ and Cr$_2$Ge$_2$Te$_6$,\cite{McGuire,Huang,Seyler,Fei,Gong,Lin,Zhuang} however the Curie temperature $T_c$ is significantly lowered when reducing to monolayer or few layers. The corresponding $T_c$ in monolayer and bilayer was only $\sim$ 45 K for CrI$_3$ and 25 K in Cr$_2$Ge$_2$Te$_6$, respectively. Therefore, high-$T_c$ ferromagnets with strong anisotropy are highly desirable.

Binary chromium tellurides Cr$_{1-x}$Te are such promising candidates with high-$T_c$ ranging from 170 to 360 K depending on different Cr vacancies.\cite{Herbert,Street,Akram,Huang1,Huang2} The carriers of magnetism in CrI$_3$ and Cr$_2$Ge$_2$Te$_6$ are Cr$^{3+}$ ions with spin $S = 3/2$ which are octahedrally coordinated by Te ligands and form a honeycomb magnetic lattice in the $ab$ plane. Cr$_{1-x}$Te crystallizes in a defective NiAs structure, showing similarities in crystal structure, magnetic ion, and magnetic ordering. The mechanism of FM in this system could be attributed to the dominant FM superexchange coupling between half-filled Cr $t_{2g}$ and empty $e_g$ states via Te $p$ orbitals, against the AFM direct exchange interactions of Cr $t_{2g}$ states. The saturation magnetization in Cr$_{1-x}$Te is observed as $1.7\sim2.5$ $\mu_B$, which is much smaller than the expected moment value of Cr using ionic model due to the spin canting, itinerant nature of the $d$ electrons and the existence of mixed valence Cr.\cite{Andresen} The electronic structure of Cr$_{1-x}$Te has further been studied by photoemission spectroscopy,\cite{Shimada} showing the importance of electron-correlation effect. Here we focus on trigonal Cr$_5$Te$_8$ with $x = 0.375$, which has a higher $T_c$ of 230 K compared with monoclinic space group.\cite{Lukoschus,Xiao,XiaoH,YuLIU,YL0} Strong magnetic anisotropy with out-of-plane easy $c$ axis has been confirmed.\cite{Akram} Critical behavior analysis of trigonal Cr$_5$Te$_8$ illustrates that the ferromagnetic state below $T_c$ conforms to three-dimensional (3D) Ising model,\cite{YuLIU} holding a high potential for establishing long-range magnetic order when thinned down to 2D limit. Anisotropic anomalous Hall effect (AHE) was recently observed in trigonal Cr$_5$Te$_8$.\cite{YL0,Wang} Calculations of band structure is desired to explain its origin, calling for precise structural information.

Magnetocaloric effect (MCE) in the FM vdW materials provides important insight into the magnetic properties. The magnetocrystalline anisotropy constant $K_u$ is found to be larger for Cr$_2$Si$_2$Te$_6$ when compared to Cr$_2$Ge$_2$Te$_6$, resulting in larger rotational magnetic entropy change at $T_c$.\cite{yuliu} CrI$_3$ also exhibits anisotropic $-\Delta S_M^{max}$ with values of 4.24 and 2.68 J kg$^{-1}$ K$^{-1}$ at 5 T for $\mathbf{H\parallel c}$ and $\mathbf{H\parallel ab}$, respectively.\cite{YULIU} Intriguingly, there is an anisotropic magnetic anomaly $T^*$ just below $T_c$ in low fields;\cite{McGuire,YULIU} the origin of $T^*$ is still unknown.

In this work we investigate the anisotropy of magnetic and magnetocaloric properties in trigonal Cr$_5$Te$_8$ single crystal. A satellite transition $T^*$ below $T_c$ was observed, at which the huge magnetic anisotropy emerges. We assign its origin to the interplay among magnetocrystalline anisotropy, temperature and magnetic field. Whereas the anisotropic magnetic entropy change $\Delta S_M$ in Cr$_5$Te$_8$ is small, the magnetocrystalline anisotropy constant $K_u$ is found to be temperature-dependent and comparable with CrI$_3$ and Cr$_2$Ge$_2$Te$_6$. As a result of high-$T_c$ in Cr$_5$Te$_8$, it would be of interest to design low-dimensional FM heterostructures.

\section{EXPERIMENTAL DETAILS}

Plate-like single crystals of Cr$_5$Te$_8$ were fabricated by the self-flux method and the surface is the $ab$ plane.\cite{YuLIU,YL0} The structure was characterized by powder x-ray diffraction (XRD) in the transmission mode at 28-D-1 beamline of the National Synchrotron Light Source II (NSLS II) at Brookhaven National Laboratory (BNL). Data were collected using a 0.5 mm$^2$ beam with wavelength $\lambda \sim$ 0.1668 {\AA}. A Perkin Elmer 2D detector (200 $\times$ 200 microns) was placed orthogonal to the beam path 990 mm away from the sample. The x-ray absorption spectroscopy measurement was performed at 8-ID beamline of NSLS II (BNL) in fluorescence mode. The x-ray absorption near edge structure (XANES) and the extended x-ray absorption fine structure (EXAFS) spectra were processed using the Athena software package. The extracted EXAFS signal, $\chi(k)$, was weighed by $k^2$ to emphasize the high-energy oscillation and then Fourier-transformed in a $k$ range from 2 to 12 {\AA}$^{-1}$ to analyze the data in $R$ space. The magnetization data as a function of temperature and field were collected using Quantum Design MPMS-XL5 system in temperature range 200 $\sim$ 280 K with a step of 4 K on cleaved crystals in order to remove surface contamination of residual Te flux droplets on the surface.

\section{RESULTS AND DISCUSSION}

\begin{table}
\caption{\label{tab}Structural parameters for Cr$_5$Te$_8$ obtained from synchrotron powder XRD at room temperature.}
\begin{ruledtabular}
\begin{tabular}{lllllll}
  \multicolumn{3}{c}{Chemical formula} &\multicolumn{4}{c}{Cr$_5$Te$_8$}\\
  \multicolumn{3}{c}{Space grroup} &\multicolumn{4}{c}{$P\bar{3}m1$}\\
  \multicolumn{3}{c}{$a$ ({\AA})} &\multicolumn{4}{c}{7.7951}\\
  \multicolumn{3}{c}{$c$ ({\AA})} &\multicolumn{4}{c}{11.9766}\\
  \hline
  atom & site & $x$ & $y$ & $z$ & Occ. & U$_{iso}$ ({\AA}$^2$) \\
  \hline
  Cr & Cr1 & 0 & 0 & 0 & 0.87 & 0.0019 \\
  Cr & Cr2 & 0.4920 & 0.5080 & 0.2515 & 1 & 0.0019 \\
  Cr & Cr3 & 0 & 0 & 0.2747 & 1 & 0.0019 \\
  Cr & Cr4 & 0.5 & 0 & 0.5 & 0.264 & 0.0019 \\
  Te & Te1 & 0.3333 & 0.6667 & 0.1164 & 1 & 0.00468 \\
  Te & Te2 & 0.3333 & 0.6667 & 0.6217 & 1 & 0.00468 \\
  Te & Te3 & 0.1652 & 0.8348 & 0.3816 & 1 & 0.00468 \\
  Te & Te4 & 0.8285 & 0.1715 & 0.1301 & 1 & 0.00468
\end{tabular}
\end{ruledtabular}
\end{table}

\begin{figure}
\centerline{\includegraphics[scale=0.8]{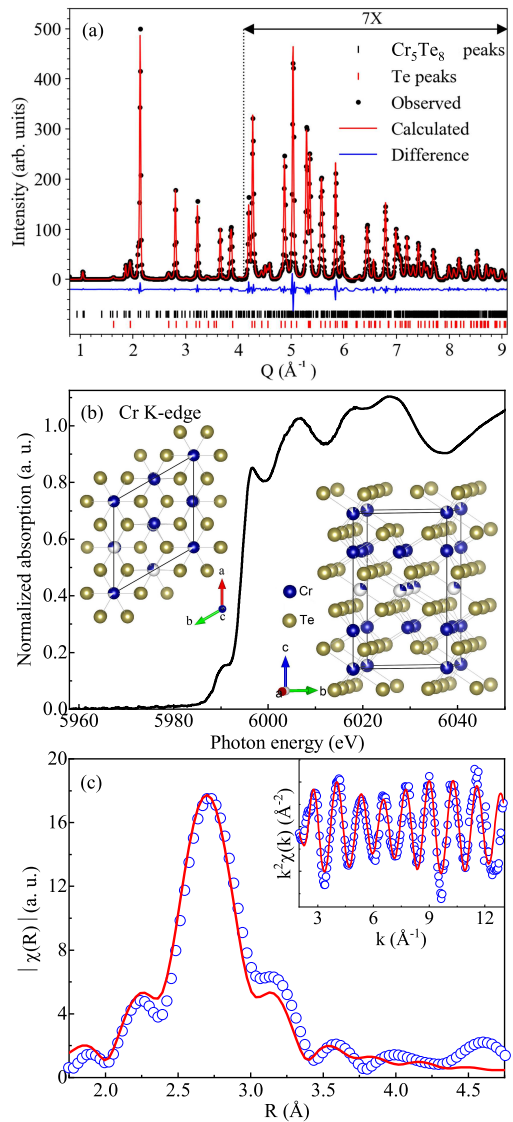}}
\caption{(Color online). (a) Refinement of synchrotron x-ray diffraction data of Cr$_5$Te$_8$: data (solid symbols), structural model (red solid line), difference curve (blue solid line, offset for clarity). Vertical tickmarks denote reflections in the main phase (black, top row) and the secondary Te impurity (red, bottom row). (b) Normalized Cr K-edge XANES spectra and (c) Fourier transform magnitudes of EXAFS data of Cr$_5$Te$_8$ taken at room temperature. The experimental data are shown as blue symbols alongside the model fit plotted as red line. The inset in (c) shows the corresponding EXAFS oscillation with the model fit.}
\label{1}
\end{figure}

\begin{figure}
\centerline{\includegraphics[scale=1]{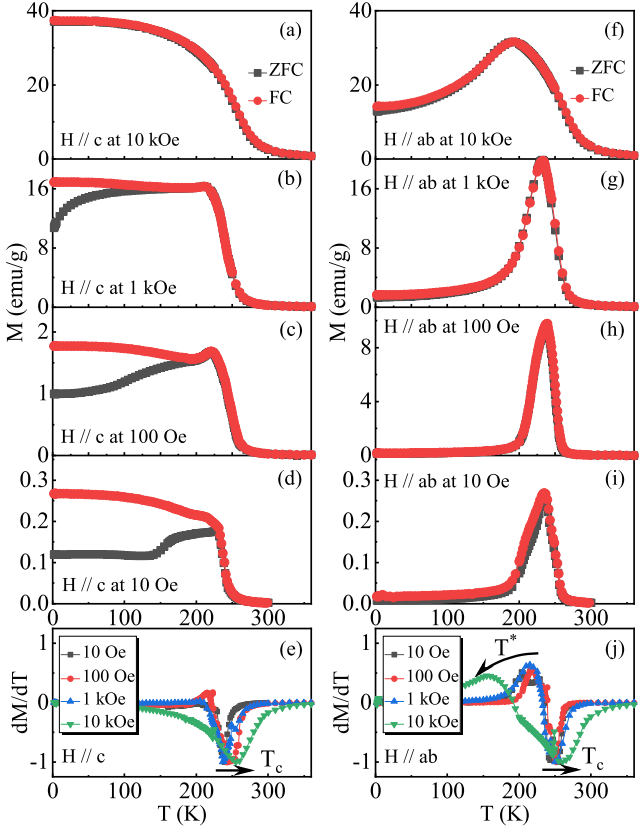}}
\caption{(Color online). (a)-(d) Temperature dependence of zero field cooling (ZFC) and field cooling (FC) magnetization of Cr$_5$Te$_8$ measured at indicated fields applied along the $c$ axis. (e) Temperature-dependent dM/dT derived from FC curves with $\mathbf{H\parallel c}$. (f)-(i) Temperature dependence of ZFC and FC magnetization of Cr$_5$Te$_8$ measured at indicated fields applied in the $ab$ plane. (j) Temperature-dependent dM/dT derived from FC curves with $\mathbf{H\parallel ab}$.}
\label{2}
\end{figure}

Rietveld powder diffraction analysis was carried out on data obtained from the raw 2D diffraction data integrated and converted to intensity versus $Q$ using the Fit2d software where $Q = 4\pi sin\theta / \lambda$ is the magnitude of the scattering vector.\cite{Hammersley} The refinement was performed using GSAS-II modeling suite.\cite{Toby} Figure 1(a) shows the refinement result (goodness of fit $\sim 4\%$) of Cr$_5$Te$_8$ (0.885 by weight) and residual Te flux impurity (0.115 by weight). The crystal structure of trigonal Cr$_5$Te$_8$ can be well refined in the $P\bar{3}m1$ space group with four crystallographically different sites for both Cr and Te atoms, leading to the formation of a five-layer superstructure of the CdI$_2$ type.\cite{Bensch} The detailed structural parameters are summarized in Table I. The superstructure is composed of three types of Cr layers with a stacking sequence abcba. Cr2 and Cr3 occupy the vacancy-free atomic layers, while Cr1 and Cr4 distribute in the metal-deficient layers [inset in Fig. 1(b)]. The average Cr-Te bond distances in vacancy-free layers range from 2.701 to 2.805 {\AA}, match well with the reported values for Cr$_2$(Si,Ge)$_2$Te$_6$ (2.741 - 2.803 {\AA}) and CrTe$_3$ (2.700 - 2.755 {\AA}).\cite{Bensch} However, the Cr-Te bond distances in metal-deficient layers are shorter with values of 2.673 - 2.681 {\AA}. The shortest Te-Te interatomic separations range from 3.664 to 3.779 {\AA}, significantly shorter than the sum of their anionic radii 4.2 {\AA}, indicative of weak bonding interactions with a quasi-2D character.

Figures 1(b) and 1(c) show the normalized Cr K-edge XANES spectra and Fourier transform magnitudes of EXAFS spectra of Cr$_5$Te$_8$, respectively. The XANES spectra is close to that of Cr$_2$Te$_3$ with Cr$^{3+}$ state.\cite{Ofuchi} The prepeak feature is due to a direct quadrupole transition to unoccupied $3d$ states that are hybridized with Te $p$ orbitals. In the single-scattering approximation, the EXAFS could be described by the following equation:\cite{Prins}
\begin{align*}
\chi(k) = \sum_i\frac{N_iS_0^2}{kR_i^2}f_i(k,R_i)e^{-\frac{2R_i}{\lambda}}e^{-2k^2\sigma_i^2}sin[2kR_i+\delta_i(k)],
\end{align*}
where $N_i$ is the number of neighbouring atoms at a distance $R_i$ from the photoabsorbing atom. $S_0^2$ is the passive electrons reduction factor, $f_i(k, R_i)$ is the backscattering amplitude, $\lambda$ is the photoelectron mean free path, $\delta_i$ is the phase shift of the photoelectrons, and $\sigma_i^2$ is the correlated Debye-Waller factor measuring the mean square relative displacement of the photoabsorber-backscatter pairs. The corrected main peak around $R \sim 2.7$ {\AA} in Fig. 1(b) corresponds to the Cr-Te bond distances [Cr2-Te1: 2.70(1) {\AA}, Cr2-Te4: 2.74(2) {\AA}, Cr4-Te2: 2.683(1) {\AA}] extracted from the model fit in the range 2 {\AA} to 3 {\AA}, only slightly modified when compared with the average bond distances. The peaks in high R range are due to longer Cr-Cr (3.01 {\AA} $\sim$ 3.29 {\AA}) and Te-Te (3.66 {\AA} $\sim$ 3.78 {\AA}) bond distances and multiple scattering effects.

\begin{figure}
\centerline{\includegraphics[scale=1]{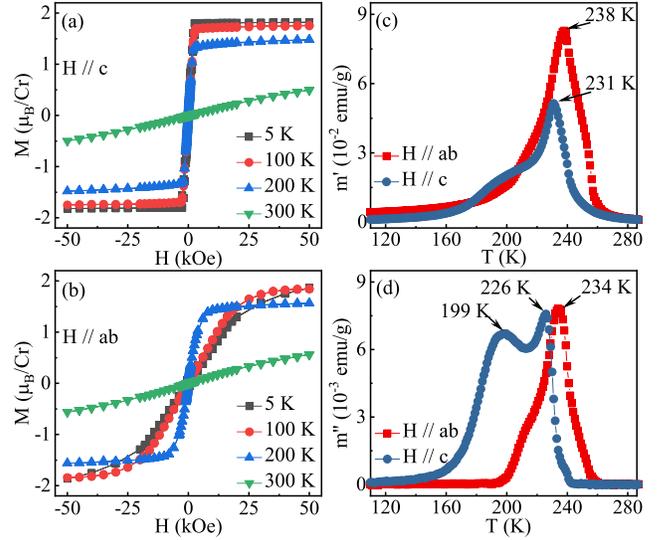}}
\caption{(Color online). Magnetization as function of field at various temperatures for (a) $\mathbf{H\parallel c}$ and (b) $\mathbf{H\parallel ab}$, respectively. Ac susceptibility real part (c) $m^\prime(T)$ and (d) imaginary part $m^{\prime\prime}(T)$ as a function of temperature measured with oscillated ac field of 3.8 Oe and frequency of 499 Hz applied in the $ab$ plane and along the $c$ axis, respectively.}
\label{3}
\end{figure}

Temperature dependence of zero field cooling (ZFC) and field cooling (FC) magnetization $M(T)$ taken at various magnetic fields for Cr$_5$Te$_8$ is presented in Fig. 2. We observe an interesting anisotropic magnetic response for fields applied along the $c$ axis and in the $ab$ plane, respectively. In Fig. 2(a), for $\mathbf{H\parallel c}$ at 10 kOe, a typical increase in $M(T)$ on cooling from high temperature corresponds well to the reported paramagnetic (PM) to FM transition.\cite{Lukoschus} An appreciable thermomagnetic irreversibility occurs between the ZFC and FC curves in lower fields [Figs. 2(b)-2(d)], in the magnetically ordered state, indicating magnetocrystalline anisotropy. On the other hand, for $\mathbf{H\parallel ab}$, an anomalous peak feature is observed. The ZFC and FC magnetization in fields ranging from 10 kOe to 10 Oe are virtually overlapping in the whole temperature range [Figs. 2(f)-2(i)]. A similar magnetization downturn below $T_c$ is also seen for Cr$_2$(Si,Ge)$_2$Te$_6$ and Cr(Br,I)$_3$,\cite{Casto,Richter} at which the strong magnetic anisotropy emerges. Since the ZFC curve might be influenced by the domain wall motion, we summarize the field dependence of $T_c$ (the minimum of $dM/dT$) and $T^*$ (the maximum of $dM/dT$) from FC magnetization. As shown in Figs. 2(e) and 2(j), both $T_c$ and $T^*$ are field-dependent but with an opposite tendency, i.e., with increasing field the $T_c$ increases whereas the $T^*$ gradually decreases.

\begin{figure}
\centerline{\includegraphics[scale=1]{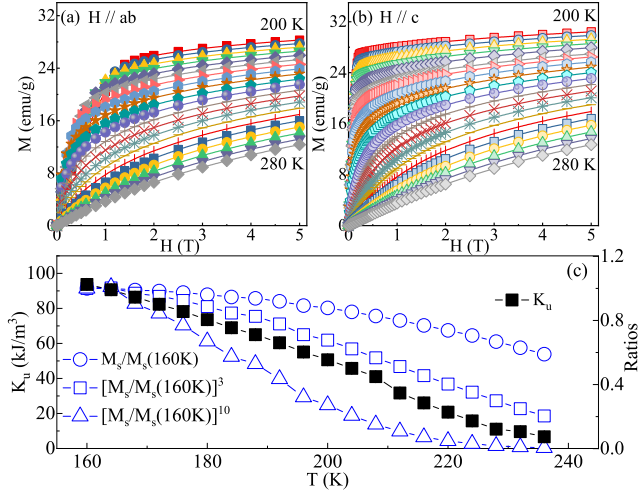}}
\caption{(Color online). Typical initial isothermal magnetization curves from $T$ = 200 K to 280 K with temperature step of 4 K measured (a) in the $ab$ plane and (b) along the $c$ axis, respectively. (c) Temperature-dependent anisotropy constant $K_u$ (left axis) and the ratios of [$M_s/M_s(160 K)$]$^{n(n+1)/2}$ with $n$ = 1, 2, and 4 (right axis).}
\label{4}
\end{figure}

Isothermal magnetization $M(H)$ for $\mathbf{H\parallel c}$ exhibits a clear FM behavior at low temperatures and a linear PM behavior with small slope at 300 K [Fig. 3(a)]. The mild hysteresis for both orientations indicates the behavior expected for a soft ferromagnet. The observed saturation magnetization of $M_s \approx$ 1.8 $\mu_B$/Cr for $\mathbf{H\parallel c}$ at 5 K which is smaller when compared to the expected value for Cr free ion (3 $\mu_B$/Cr), indicating weak itinerant nature of Cr or canted FM in Cr$_5$Te$_8$.\cite{YuLIU} The saturation field $H_s$ could be derived from the $x$ component of the intercept of two linear fits, one being a fit to the saturated regime at high fields and one being a fit of the unsaturated linear regime at low fields. As we can see, the magnetization is much easier to saturate for $\mathbf{H\parallel c}$ [Fig. 3(a)] than for $\mathbf{H\parallel ab}$ [Fig. 3(b)], indicating that the $c$ axis is the easy magnetization direction. The large disparity of $H_s$ for two orientations can be explained by easy (hard) natures of the domain wall motion along different $c$ (ab) axes. Figures 3(c) and 3(d) present the temperature dependence of ZFC ac susceptibility measured with oscillated ac field of 3.8 Oe and frequency $f$ = 499 Hz. A pronounced broad peak is observed at $T_c$ in the real part $m^\prime$, i. e. at 238 K for $\mathbf{H\parallel ab}$ and at 231 K for $\mathbf{H\parallel c}$. An additional anomaly occurs just below $T_c$, as seen in an obvious hump in the imaginary part $m^{\prime\prime}$ at 199 K for $\mathbf{H\parallel c}$, further confirming the anisotropic magnetic response.

\begin{figure}
\centerline{\includegraphics[scale=1]{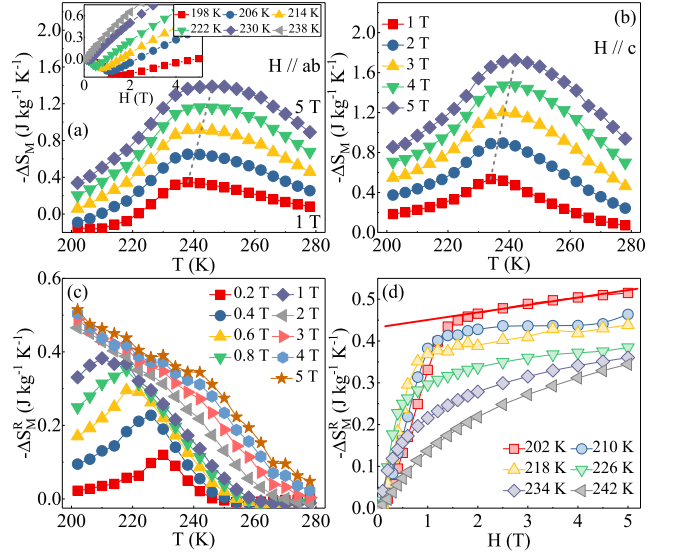}}
\caption{(Color online). Temperature dependence of magnetic entropy change $-\Delta S_M$ obtained from magnetization measurement at various magnetic field changes (a) in the $ab$ plane and (b) along the $c$ axis, respectively. Inset shows the field dependence of $-\Delta S_M$ at low temperature. The rotational magnetic entropy change $-\Delta S_M^R$ as a function of (c) temperature and (d) field.}
\label{5}
\end{figure}

Using the Stoner-Wolfarth model a value for the magnetocrystalline anisotropy constant $K_u$ can be estimated from the saturation regime in the isothermal magnetization curve. Figures 4(a) and 4(b) present the magnetization isotherms with field up to 5 T in the temperature range from 200 to 280 K for both $\mathbf{H\parallel ab}$ and $\mathbf{H\parallel c}$. When $\mathbf{H\parallel ab}$, where the anisotropy becomes maximal, the saturation field $H_{s}$ is associated with the magnetocrystalline anisotropy constant $K_u$ and the saturation magnetization $M_s$ with $\mu_0$ being the vacuum permeability: $2K_u/M_s = \mu_0H_{s}$.\cite{Cullity} Figure 4(c) exhibits the temperature dependence of derived $K_u$ for Cr$_5$Te$_8$, which is about 94 kJ cm$^{-3}$ at $T$ = 160 K. The $K_u$ in Cr$_5$Te$_8$ is larger than those for Cr$_2$(Si,Ge)$_2$Te$_6$ and CrBr$_3$,\cite{yuliu,Richter} which most likely results in a more apparent magnetization downturn anomaly in Cr$_5$Te$_8$ [Figs. 2(f)-2(i)]. Here we propose that the origin of this downturn in the magnetization curve for $\mathbf{H\parallel ab}$ is a continuous reorientation of the magnetization direction as a result of an interplay between the magnetocrytalline anisotropy, field and temperature. The magnetocrytalline anisotropy favors a magnetization direction perpendicular to the $ab$ plane, which for $\mathbf{H\parallel ab}$ the field wants to align the magnetization direction parallel to the field. Assuming an external field $H_1$ is higher than the anisotropy field $H^*$ at temperature $T_1$ below $T^*$, the magnetization vector is aligned along the external field direction. By reducing the external field to a value $H_2$ below $H^*$ at fixed $T_1$, a tilting of the magnetization vector towards the easy $c$ axis will be achieved. The tilting in turn leads to a reduction of the magnetization component parallel to the field. This is like the spin-flop mechanism that usually is observed in an AFM ordered state. Such feature is also presented in the field-dependent magnetization in the hard $ab$ plane [Fig. 3(b)] at 5 and 100 K.

\begin{figure}
\centerline{\includegraphics[scale=1]{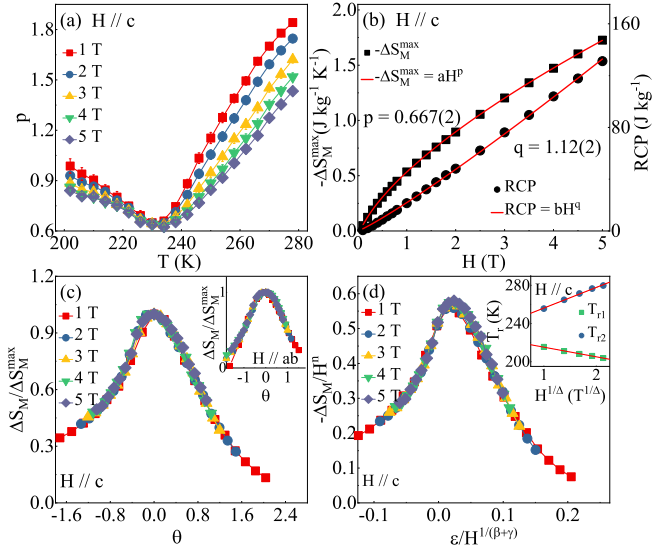}}
\caption{(Color online). (a) Temperature dependence of $p$ in various fields. (b) Field dependence of the maximum magnetic entropy change $-\Delta S_M^{max}$ and the relative cooling power RCP with power law fitting in red solid lines. (c) The normalized $\Delta S_M$ as a function of the rescaled temperature $\theta$ with out-of-plane field and in-plane field (inset). (d) Scaling plot of $\Delta S_M$ based on the critical exponents $\beta$ = 0.315 and $\gamma$ = 1.81.\cite{YuLIU} Inset shows $T_r$ vs $H^{1/\Delta}$ with $\Delta = \beta + \gamma$.}
\label{6}
\end{figure}

Furthermore, the magnetocrytalline anisotropy $K_u$ for Cr$_5$Te$_8$ is found to be temperature-dependent [Fig. 4(c)], gradually increasing with decreasing temperature. This tendency arises solely from a large number of local spin clusters fluctuating randomly around the macroscopic magnetization vector and activated by a nonzero thermal energy.\cite{Zener,Carr} In a simple classical theory, $\langle K^{(n)}\rangle \propto M_s^{n(n+1)/2}$, where $\langle K^{(n)}\rangle$ is the anisotropy expectation value for the $n^{th}$ power angular function,\cite{Zener,Carr} in the case of uniaxial anisotropy $n$ = 2 and of cubic anisotropy $n$ = 4, leading to an exponent of 3 and 10, respectively. The temperature-dependent $M_s/M_s(160 K)$, $[M_s/M_s(160 K)]^3$, and $[M_s/M_s(160 K)]^{10}$ are also plotted. The comparison in Fig. 5(c) points to the predominant uniaxial anisotropy with some deviation to cubic anisotropy in Cr$_5$Te$_8$. As discussed above, reducing the external field from $H_1$ passing $H^*$ to $H_2$ at $T_1$ ($< T^*$) results in a tilting of the magnetization vector towards the easy $c$ axis. The observed temperature-dependent $K_u$ gives a reduction of magnetic anisotropy when increasing temperature to $T_2$ ($> T^*$). Then the alignment along the magnetic easy axis becomes less favorable upon warming, which leads to a stronger tilting of the magnetization vector towards the $ab$ plane and an increased experimentally observed $ab$-component magnetization. Therefore, the spin configuration of Cr$_5$Te$_8$ below $T_c$ is supposed to be an out-of-plane FM order with in-plane AFM components, calling for further neutron scattering measurement. This canted FM configuration might arise from the ordered Cr-vacancies induced modification of local structure and indicates competing exchange interactions exist in Cr$_5$Te$_8$.

To further characterize the anisotropic magnetic properties in Cr$_5$Te$_8$, we investigate the anisotropic magnetic entropy change:
\begin{align*}
\Delta S_M(T,H) = \int_0^H \left(\frac{\partial S}{\partial H}\right)_TdH = \int_0^H \left(\frac{\partial M}{\partial T}\right)_HdH,
\end{align*}
where $\left(\frac{\partial S}{\partial H}\right)_T$ = $\left(\frac{\partial M}{\partial T}\right)_H$ is based on Maxwell's relation.\cite{Pecharsky} In the case of magnetization measured at small discrete field and temperature intervals,
\begin{align*}
\Delta S_M(T_i,H) = \frac{\int_0^HM(T_i,H)dH-\int_0^HM(T_{i+1},H)dH}{T_i-T_{i+1}}.
\end{align*}
Figures 5(a) and 5(b) show the calculated $-\Delta S_M(T,H)$ as a function of temperature in selected fields up to 5 T with a step of 1 T applied in the $ab$ plane and along the $c$ axis, respectively. All the $-\Delta S_M(T,H)$ curves feature a broad peak in the vicinity of $T_c$. The value of $-\Delta S_M$ increases monotonically with increasing field; the maximal value reaches 1.39 J kg$^{-1}$ K$^{-1}$ in the $ab$ plane and 1.73 J kg$^{-1}$ K$^{-1}$ along the $c$ axis, respectively, with field change of 5 T. Most importantly, the values of $-\Delta S_M$ for the $ab$ plane are negative at low temperatures in low fields, however, all the values are positive along the $c$ axis. The field dependence of $-\Delta S_M(T,H)$ at low temperature [inset in Fig. 5(a)] shows a clear sign change for $\mathbf{H\parallel ab}$. This behavior is similar with that in CrI$_3$,\cite{YULIU} in which the magnetic anisotropy originates mainly from the superexchange interaction.\cite{Lado} The calculated magnetocrytalline anisotropy decreases with increasing temperature [Fig. 4(c)], whereas the magnetization may exhibit opposite behavior. At low fields, the magnetization at higher temperature could be larger than that at lower temperature [Figs. 2(f)-(i) and 3(b)], which gives a negative $-\Delta S_M$. The rotational magnetic entropy change $\Delta S_M^R$ can be calculated as $\Delta S_M^R(T,H) = \Delta S_M(T,H_c)-\Delta S_M(T,H_{ab})$. As shown in Fig. 5(c), the temperature dependence of $-\Delta S_M^R(T,H)$ also features a peak at $T_c$ with the field change of 0.2 T, however, the temperature of the maximum $-\Delta S_M^R(T,H)$ moves away from $T_c$ to lower temperature with increasing field. When $T \leq T_c$, the field dependence of $-\Delta S_M^R(T,H)$ increases rapidly at low field and changes slowly at high field [Fig. 5(d)]. This possibly reflects the anisotropy field in Cr$_5$Te$_8$, which decreases with increasing temperature and gradually disappears when $T \geq T_c$. The obtained $-\Delta S_M$ of Cr$_5$Te$_8$ is significantly smaller than those of well-known magnetic refrigerating materials, such as Gd$_5$Si$_2$Ge$_2$, LaF$_{13-x}$Si$_x$, and MnP$_{1-x}$Si$_x$,\cite{GschneidnerJr} however, comparable with those of Cr(Br,I)$_3$ and Cr$_2$(Si,Ge)$_2$Te$_6$.\cite{yuliu,Xiaoyun,YS}

For a material displaying a second-order transition,\cite{Oes} the field-dependent maximum magnetic entropy change should be $-\Delta S_M^{max} \propto H^p$.\cite{VFranco} Another important parameter is the relative cooling power $RCP = -\Delta S_M^{max} \times \delta T_{FWHM}$.\cite{Gschneidner} The $-\Delta S_M^{max}$ is the maximum entropy change near $T_c$ and $\delta T_{FWHM}$ is the full-width at half maximum of $-\Delta S_M$. The RCP also depends on the field with $RCP \propto H^q$. Figure 6(a) displays the temperature dependence of $p(T)$ in various out-of-plane fields. At low temperatures, well below $T_c$, $p$ has a value which tends to 1. On the other side, well above $T_c$, $p$ approaches 2 as a consequence of the Curie-Weiss law. At $T = T_c$, $p$ reaches a minimum. Figure 6(b) summarized the field dependence of $-\Delta S_M^{max}$ and RCP. The RCP is calculated as 131.2 J kg$^{-1}$ with out-of-plane field change of 5 T for Cr$_5$Te$_8$, which is comparable with that of CrI$_3$ (122.6 J kg$^{-1}$ at 5 T).\cite{YULIU} Fitting of the $-\Delta S_M^{max}$ and RCP gives $p = 0.667(2)$ and $q = 1.12(2)$.

Scaling analysis of $-\Delta S_M$ can be built by normalizing all the $-\Delta S_M$ curves against the respective maximum $-\Delta S_M^{max}$, namely, $\Delta S_M/\Delta S_M^{max}$ by rescaling the reduced temperature $\theta_\pm$ as defined in the following equations,\cite{Franco}
\begin{align*}
\theta_- = (T_{peak}-T)/(T_{r1}-T_{peak}), T<T_{peak},
\end{align*}
\begin{align*}
\theta_+ = (T-T_{peak})/(T_{r2}-T_{peak}), T>T_{peak},
\end{align*}
where $T_{r1}$ and $T_{r2}$ are the temperatures of two reference points that corresponds to $\Delta S_M(T_{r1},T_{r2}) = \frac{1}{2}\Delta S_M^{max}$. Following this method, all the $-\Delta S_M(T,H)$ curves in various fields collapse into a single curve [Fig. 6(c) and inset]. The values of $T_{r1}$ and $T_{r2}$ depend on $H^{1/\Delta}$ with $\Delta = \beta + \gamma$ [inset in Fig. 6(d)].\cite{Franco} In the phase transition region, the scaling analysis of $-\Delta S_M$ can also be expressed as $-\Delta S_M/a_M = H^nf(\varepsilon/H^{1/\Delta})$, where $a_M = T_c^{-1}A^{\delta+1}B$ with A and B representing the critical amplitudes as in $M_s(T) = A(-\varepsilon)^\beta$ and $H = BM^\delta$, respectively, and $f(x)$ is the scaling function.\cite{Su} If the critical exponents are appropriately chosen, the $-\Delta S_M(T)$ curves should be rescaled into a single curve, consistent with normalizing all the $-\Delta S_M$ curves with two reference temperatures ($T_{r1}$ and $T_{r2}$). This is indeed seen in Fig. 6(d) by using the values of $\beta$ = 0.315 and $\gamma$ = 1.81,\cite{YuLIU} confirming that the critical exponents for Cr$_5$Te$_8$ are intrinsic and accurately estimated. As is well known, the critical exponents $\beta$ = 0.25, $\gamma$ = 1.75 and $\beta$ = 0.325, $\gamma$ = 1.24 are expected for 2D and 3D Ising models, respectively. The spatial dimension extends from $d$ = 2 to $d$ = 3 passing through $T_c$ confirms that trigonal Cr$_5$Te$_8$ features quasi-2D character. The spin dimension of Ising-type $n$ = 1 implies strong uniaxial magnetic anisotropy, which is of high interest for establishing long-range magnetism down to 2D limit.\cite{YL}

\section{CONCLUSIONS}

In summary, we have studied in detail the anisotropy of magnetic and magnetocaloric properties of trigonal Cr$_5$Te$_8$ with a high-$T_c$ of 230 K. A satellite transition $T^*$ is confirmed below $T_c$, featuring an anomalous magnetization downturn when low field applying in the hard $ab$ plane. A canted out-of-plane FM configuration with in-plane AFM components is supposed for Cr$_5$Te$_8$, suggesting the existence of competing exchange interactions. Intrinsic magnetocrystalline anisotropy in trigonal Cr$_5$Te$_8$ is established and also reflected in anisotropic magnetic entropy change. Taken its high-$T_c$ and strong anisotropy, Cr$_5$Te$_8$ is a promising material to gain further insight into low-dimensional ferromagnetism and is of high interest or nanofabrication that could led to ferromagnetic heterostructures devices.

\section*{Acknowledgements}

This work was funded by the Computation Material Science Program (Y.L. and C.P.). This research used the 28-ID-1 and 8-ID beamlines of the National Synchrotron Light Source II, a U.S. DOE Office of Science User Facility operated for the DOE Office of Science by Brookhaven National Laboratory under Contract No. DE-SC0012704.

\end{document}